


\magnification\magstep1
\parskip=\medskipamount
\hsize=6.0 truein
\vsize=8.2 truein
\hoffset=.2 truein
\voffset=0.4truein
\baselineskip=13pt
\tolerance 500


\font\titlefont=cmbx12
 at 10 truept
\font\authorfont=cmcsc10
\font\addressfont=cmsl10 at 10 truept
\font\smallbf=cmbx10 at 10 truept


\outer\def\beginsection#1\par{\vskip0pt plus.2\vsize\penalty-150
\vskip0pt plus-.2\vsize\vskip1.2truecm\vskip\parskip
\message{#1}\leftline{\bf#1}\nobreak\smallskip\noindent}


\newdimen\itemindent \itemindent=13pt
\def\textindent#1{\parindent=\itemindent\let\par=\resetpar%
\indent\llap{#1\enspace}\ignorespaces}

\let\oldpar=\par
\def\resetpar{\oldpar\parindent=0pt\let\par=\oldpar}

\font\ninerm=cmr9 \font\ninesy=cmsy9
\font\eightrm=cmr8 \font\sixrm=cmr6
\font\eighti=cmmi8 \font\sixi=cmmi6
\font\eightsy=cmsy8 \font\sixsy=cmsy6
\font\eightbf=cmbx8 \font\sixbf=cmbx6
\font\eightit=cmti8
\def\eightpoint{\def\rm{\fam0\eightrm}
  \textfont0=\eightrm \scriptfont0=\sixrm \scriptscriptfont0=\fiverm
  \textfont1=\eighti  \scriptfont1=\sixi  \scriptscriptfont1=\fivei
  \textfont2=\eightsy \scriptfont2=\sixsy \scriptscriptfont2=\fivesy
  \textfont3=\tenex   \scriptfont3=\tenex \scriptscriptfont3=\tenex
  \textfont\itfam=\eightit  \def\it{\fam\itfam\eightit}%
  \textfont\bffam=\eightbf  \scriptfont\bffam=\sixbf
  \scriptscriptfont\bffam=\fivebf  \def\bf{\fam\bffam\eightbf}%
  \normalbaselineskip=9pt
  \setbox\strutbox=\hbox{\vrule height7pt depth2pt width0pt}%
  \let\big=\eightbig \normalbaselines\rm}
\catcode`@=11 %
\def\eightbig#1{{\hbox{$\textfont0=\ninerm\textfont2=\ninesy
  \left#1\vbox to6.5pt{}\right.\n@space$}}}
\def\vfootnote#1{\insert\footins\bgroup\eightpoint
  \interlinepenalty=\interfootnotelinepenalty
  \splittopskip=\ht\strutbox %
  \splitmaxdepth=\dp\strutbox %
  \leftskip=0pt \rightskip=0pt \spaceskip=0pt \xspaceskip=0pt
  \textindent{#1}\footstrut\futurelet\next\fo@t}
\catcode`@=12 %


\def\mapright#1{\smash{
  \mathop{\longrightarrow}\limits^{#1}}}
\def\mapdown#1{\Big\downarrow
  \rlap{$\vcenter{\hbox{$\scriptstyle#1$}}$}}
\def\diag{
  \def\normalbaselines{\baselineskip2pt \lineskip3pt
    \lineskiplimit3pt}
  \matrix}

\def\maprights#1{\smash{
  \mathop{\rightarrow}\limits^{#1}}}
\def\mapdowns#1{\big\updownarrow
  \rlap{$\vcenter{\hbox{$\scriptstyle#1$}}$}}


\def\b#1{\hbox{$\bar#1$}}
\def\h#1{\hbox{$\hat#1$}}
\def\S{\bar\Sigma}
\def\shalf{\hbox{${\textstyle{1\over 2}}$}}

\def\G{\hbox{$\cal G$}}
\def\phii{\phi_{\infty}}
\def\Si{\Sigma}
\def\Sib{\bar{\Sigma}}
\def\S{\hbox{$\cal S$}}
\def\I{\hbox{$\cal I$}}
\def\R{\hbox{$\cal R$}}
\def\D{\hbox{$\cal D$}}
\def\Re{I\!\! R}

\tolerance=500


\rightline{CGPG-94/10-}
\rightline{gr-qc/9410042}
\bigskip
{\baselineskip=24 truept
\titlefont
\centerline{ASYMPTOTIC SYMMETRY GROUPS OF}
\centerline{LONG-RANGED GAUGE CONFIGURATIONS}
}

\vskip 1.1 truecm plus .3 truecm minus .2 truecm

\centerline{\authorfont Domenico Giulini\footnote*{
On leave from: Fakult\"at f\"ur Physik, Universit\"at Freiburg,
Hermann-Herder Stra\ss e 3, D-79104 Freiburg i.Br., Germany
}}
\vskip 2 truemm
{\baselineskip=12truept
\addressfont
\centerline{Center for Gravitational Physics and Geometry}
\centerline{The Pennsylvania State University}
\centerline{104 Davey Laboratory, University Park, PA 16802-6300, USA}
}
\vskip 1.5 truecm plus .3 truecm minus .2 truecm

\centerline{\smallbf Abstract}
\vskip 1 truemm
{\baselineskip=12truept
\parindent=0pt

{\eightpoint
We make some general remarks on long-ranged configurations in gauge
or diffeomorphism invariant theories where the fields are allowed to
assume some non vanishing values at spatial infinity. In this case the
Gauss constraint only eliminates those gauge degrees of freedom
which lie in the connected component of asymptotically trivial
gauge transformations. This implies that proper physical
symmetries arise either from gauge transformations that reach to
infinity or those that are asymptotically trivial but do not
lie in the connected component of transformations within that class.
The latter transformations form a discrete subgroup of all
symmetries whose position in the ambient group has
proven to have interesting implications. We explain this for the
dyon configuration in the $SO(3)$ Yang-Mills-Higgs theory, where we
prove that the asymptotic symmetry group is $Z_{|m|}\times \Re$
where $m$ is the monopole number.  We also discuss the application
of the general setting  to general relativity and show that
here the only implication of discrete symmetries for the continuous
part is a possible extension of the rotation group $SO(3)$ to
$SU(2)$.
\par}}

\beginsection{Introduction}

In theories with gauge or diffeomorphism invariance some of the
canonical variables do not really label physically existent
degrees of freedom. Rather, they are labels on a phase space,
$\Gamma$, whose points represent physical states in a redundant
fashion. We leave the question aside as to whether the employment
of redundant labelings is purely a matter of convenience or points
towards some deeper underlying necessity. In any case, the
Hamiltonian formulations of such theories display that fact by
presenting constraints -- usually  called Gauss constraints for
gauge theories or diffeomorphism  constraints for general
relativity\footnote*{In order to avoid too many repetitions we
shall in this section adopt the convention that gauge transformations
and Gauss constraints are collective names also including diffeomorphism
and the diffeomorphism constraints respectively.} -- which define
the constraint surface $\b\Gamma\subset\Gamma$ (their zero level set) and
whose Hamiltonian flows connect points which label the {\it same}
physical state. Through each point of $\b \Gamma$ passes exactly
one of the orbits generated by the Gauss constraints.
The program of Hamiltonian reduction (see e.g. [1]) now advises to construct
the so-called reduced phase space,
$\h \Gamma$, which is just the space of gauge orbits in $\b \Gamma$.
It would then furnish a faithful phase space where any two
different points label different physical states.
This can be interpreted as saying that there are sufficiently
many physical observables to separate any two points in
$\h \Gamma$. Clearly this cannot be the case on $\Gamma$ or
$\b \Gamma$ if physical observables are required to (Poisson)
commute with the Gauss constraints. Unfortunately
the construction of $\h \Gamma$ is forbiddingly difficult
in many cases of physical interest. This makes it a working
necessity to employ the redundant label space $\Gamma$
with unsolved constraints.

In many cases, including those mentioned above, the redundant
phase space has the structure of an $\R$-principal bundle, where
$\R$ now denotes the
group that is generated by the Gauss constraints. We call it
the {\it redundancy group}. A crucial remark is that this
group is generally only a normal proper subgroup of the group
of all admissible gauge transformations which we call the
{\it invariance group}, denoted by $\I$. Both of these groups
have the property of mapping solution curves to the Hamiltonian
equations on $\Gamma$ onto solution curves. However, neither of these
deserves to be called a symmetry. To make this clear, we pretend
for the moment that we succeeded in the
construction of $\h \Gamma$. Then the quotient $\S:=\I/\R$ would act
on $\h \Gamma$ thereby mapping solution curves to solution curves.
But since $\h \Gamma$ is a faithful label space for physical states,
we may say that $\S$ maps solution curves to new, {\it physically
different} solution curves. It is this feature of being a
{\it physically} active transformation group that,
in our opinion, distinguishes a symmetry transformation from a
mere redundancy. We thus call $\S$ the {\it symmetry group}.
The relation of the three groups introduced is compactly displayed
by the following sequence of groups and homomorphisms, where at
each step the image of the `arriving' map equals the
kernel of the `departing' one. One says the sequence is exact. (The
unit $1$ denotes the trivial group.):
$$
\diag{
1&\mapright{}&\R&\mapright{i}&\I&\mapright{p}&\S&\mapright{}&1\cr}
\eqno{(1)}
$$
Here the map $i$ denotes the injective inclusion map, and $p$
the surjective quotient map from $\I$ onto $\S:=\I/\R$. In
group theoretic terms, $\I$ is an $\R$-extension of $\S$.

However, we here want to consider the situation where one works with
$\Gamma$ rather than $\h \Gamma$. In this case we are primarily given
the invariance group $\I$ and the normal subgroup $\R$ acting on
$\Gamma$ (or $\b \Gamma$). The symmetry group $\S$ then arises only as
a quotient and cannot generally be expected to also act on $\Gamma$.
This is precisely what happens in gauge theories or general relativity.
In these cases the invariance group is given the  group of all gauge
transformations on a Cauchy hypersurface. If this hypersurface is open
with one asymptotic region, as we assume, then $\I$ has to leave invariant
the boundary conditions in the asymptotic region. Those boundary
conditions may well include  those for which the fields assume nonzero and
possibly non constant values in the asymptotic region. The group $\R$ is
then given by the identity component of asymptotically trivial gauge
transformations, where the condition of asymptotic triviality refers to
certain falloff condition that must be met in order for them to be
generated by the Gauss constraint. In the following sections we will
specify the groups $\R$ and $\I$ for long ranged configurations in
Yang-Mills-Higgs theory and determine the  symmetry group $\S$.
The last section deals with general relativity. These parts of the
present work may be considered as an elaboration on some aspects
presented in less detail in [2] and [3].

\beginsection{Symmetries of SO(3)-Dyon Configurations}

We consider the standard $SO(3)$ Yang-Mills-Higgs model, with the
Higgs field $\phi$ in the adjoint representation and symmetry braking
potential $V(\phi)$. As spacetime we take ${\Re}^4$. The $SO(3)$
bundle  is topologically trivial so that gauge transformations can
be identified with $SO(3)$-valued functions.

We identify the $SO(3)$ Lie algebra with ${\Re}^3$ in the standard fashion:
$so(3)\ni\{L_{ij}\}\mapsto \{\shalf\varepsilon^{aij}L_{ij}\}=
:\{L^a\}=:\vec L$. The connection is called ${\vec A}_{\mu}$ and the
ad-covariant derivative reads $D_{\mu}=\partial_{\mu}+{\vec A}_{\mu}\times$,
where the vector product is defined as usual.
${\vec L}\cdot {\vec M}$ then denotes the standard inner product on $\Re^3$
which on the Lie algebra corresponds to $-\shalf\hbox{trace}$.

As an orientation we remind on the asymptotic behavior of the dyon
solution found by Julia and Zee [4] (see also [5] for a
comprehensive account).
$$
\eqalignno{
A^a_k(r\rightarrow\infty) \propto\ &
\varepsilon_{aki}{n^i\over r}+\alpha/r^2+O(1/r^2) &(2a)\cr
A^a_0(r\rightarrow\infty) \propto\ &
n^a+\beta/r+O(1/r)                              &(2b)\cr
\phi^a(r\rightarrow\infty) \propto\ &
n^a+\gamma/r+O(1/r)                             &(2c)\cr}
$$
where $\alpha$, $\beta$, and $\gamma$ are certain constants and $O(1/r^n)$
denotes terms with falloff faster then $1/r^n$. In the
Hamiltonian formulation the (redundant) phase space $\Gamma$ is
labeled by the gauge connection ${\vec A}_i$, the Higgs field
$\vec \phi$ and their momenta ${\vec \pi}^i$ and  $\vec \pi$
respectively. Asymptotically the Higgs field is required to
approach the so-called Higgs vacuum which is defined by $\vec \phi$
assuming values in the vacuum manifold
$S_H=\{\vec \phi\,/\,\Vert\vec \phi\Vert=a\}$
(the two-sphere of radius $a$ in the Lie algebra, where $a$ sets the
symmetry braking scale) and be covariantly constant: $D_{\mu}\vec \phi=0$.
The boundary condition for the Higgs field is thus
given by a map $\phii:S_{\infty}\rightarrow S_H$ whose degree
(also called winding number) $m$ is identified with the monopole
number. For example, the ``radial'' map (2c) has degree $m=1$.
But  since $\vec \phi$ approaches a radially independent value
$\vec\phii$ (i.e. depending only on $n^i=x^i/r$), its partial
derivatives $\partial_k\vec \phi$ must approach zero. Since the
covariant derivatives must also approach zero, ${\vec A}_k$ must
also approach zero. This is exemplified by (2), where the only
asymptotically non vanishing gauge potential is ${\vec A}_0$,
which in the canonical theory becomes the generator for gauge
transformations.

A Lie algebra valued map $\vec \Lambda$ defines an infinitesimal gauge
transformation according to
$$\eqalignno{
\delta {\vec A}_{k} & =D_k\vec \Lambda            &(3a)\cr
\delta\vec \phi   & =-\vec \Lambda\times\vec \phi &(3b)\cr}
$$
In order to preserve the asymptotic behavior of $\vec \phi$ the
asymptotically  non vanishing part must be proportional to $\vec \phi$:
$$
\vec\Lambda(r\rightarrow\infty)=
\eta \vec \phi+{\lambda (\omega)\over r}+O(1/r)
\eqno{(3c)}
$$
Inserting this into (3a) shows that $\delta A_k$ falls off faster than
$1/r$ if $\partial_k\eta$ falls off faster than $1/r$.
The phase space function that generates the transformations (3) is given
by
$$
I^{\Lambda}=\int_{{\Re}^3}d^3x\,\left\{{\vec\pi}^k\cdot D_k\vec\Lambda+
(\vec\pi\times\vec\phi)\cdot\vec\Lambda\right\}
\eqno{(4)}
$$
In contrast, the Gauss law reads
$$
\vec G = -D_k{\vec\pi}^k+\vec\pi\times\vec\phi=0
\eqno{(5)}
$$
so that
$$
\int_{{\Re}^3}d^3x\,\vec\Lambda\cdot \vec G=I^{\Lambda}+
\int_{S_{\infty}}d\omega\,({\vec\pi}^kn_k)\cdot\vec\Lambda
\eqno{(6)}
$$
This tells us that the Gauss constraint generates only those gauge
transformations for which the surface term in (6) vanishes. Since
we wish to allow a $1/r^2$ falloff for the field strength, this
means that $\Lambda$ must approach zero at infinity. The redundancy
group $\R$, which was defined to be the group generated by the Gauss
constraint, is thus seen to be given by the identity component of
all asymptotically trivial gauge transformations. Their group will
thus be called $\G_F$ ($F$ to remind on the falloff condition) and its
identity component $\G_F^0$. On the other hand, gauge transformations that
asymptotically approach rotations about the  Higgs field  are still
allowed. One says that asymptotically the gauge group is broken from
$SO(3)$ to $U(1)$, where $U(1)$ labels the rotation angle about the
Higgs field. This group of
residual gauge transformations contains as a subgroup those that
asymptotically assume a constant value in $U(1)$. These we take as our
invariance group $\I$, which we now call $\G_{\infty}$, where the
subscript $\infty$ reminds us on the  explicit dependence on the
asymptotic Higgs field. Modulo the asymptotically trivial gauge
transformations, these invariances consist of what one might call the
{\it global} $U(1)$, since $\G_{\infty}/\G_F\cong U(1)$ (throughout
we use the symbol $\cong$ to denote structural isomorphisms).
Formally we can characterize $\G_{\infty}$ by saying that the maps in
$\G_{\infty}$ extend to the one point compactification
$({\Re}^3,\infty)\cong S^3$, with $\infty$ being the point `infinity'.
This restriction does in fact not imply a loss of generality concerning
those aspects we are interested in here, as we will
briefly point out at the end of this section. In the sequel we maintain
the symbol $\S$ for the symmetry group in each case.

After these general remarks, we now wish to determine the symmetry
group $\S$. As just pointed out, the quotient $\G_{\infty}/\G_F$
is given by the group $U(1)$ whose points label the rotation value
(about the Higgs field) at infinity. The group $\G_{\infty}$ can in
fact be regarded as the principal fiber bundle
with fiber $\G_F$ and base
$U(1)$:
$$
\diag{
\G_F&\mapright{i}&\G_{\infty}\cr
&&\mapdown{p}\cr
&&U(1)\cr}
\eqno{(7)}
$$
where the projection map $p$ just evaluates the functions in $\G_{\infty}$
at $\infty$. Associated to it is the bundle obtained by taking as structure
group the group of connected components. It is obtained from (7) by taking
the quotient of the total space with respect to $\G_F^0$. We
identify $\G_F$ with the space of based maps $({\Re}^3,\infty)
\rightarrow (SO(3),e)$ where $e$ denotes the identity in $SO(3)$ and where
by $({\Re}^3, \infty)$ we mean the already mentioned $S^3$
compactification of ${\Re}^3$ with basepoint $\infty$. The
different connected components are thus labeled by the homotopy
group $\pi_3(SO(3))\cong Z$, which is generated by the standard
covering map $S^3\rightarrow \Re P^3$ of degree
two\footnote*{If one defines winding number as the degree, this
is at variance with some statements in the literature.
Let us therefore recall that the definition of degree
is given by the sum over the signs of the Jacobian determinant
for all preimage points of some regular value. There is simply no
map of odd degree from $S^3$ to $\Re P^3$.}.
We thus obtain the associated bundle
$$
\diag{
Z\cong \G_F/\G_F^0&\mapright{i}&
\G_{\infty}/\G_F^0\cong \S\quad\cr
&&\mapdown{p}\qquad\quad\cr
&&\G_{\infty}/\G_F\cong U(1)\cr}
\eqno{(8)}
$$
We wish to know how these two abelian groups, $Z$ and $U(1)$,
are combined topologically. This can be deduced by looking at the
nontrivial piece of the exact sequence for the bundle (8):
$$
\diag{
1&\maprights{}&\pi_1(\S)&\maprights{p_*}&Z&\maprights{\partial_*}&
Z&\maprights{i_*}&\S/\S^0&\maprights{}&1\cr
&&&&\mapdowns{iso}&&\mapdowns{iso}&&&\cr
&&&&\pi_1(U(1))&&\G_F/\G_F^0&&&\cr}
\eqno{(9)}
$$
Here the homomorphism  $\partial_*$ can be described as follows:
Take a loop $\gamma_t$, $t\in [0,1]$, in $U(1)$ based at the identity
$e$ whose homotopy class $[\gamma_t]$ generates $\pi_1(U(1))\cong Z$.
Let $\epsilon\in p^{-1}(e)$ be the identity of $\S$.
Now lift $\gamma_t$ to a curve ${\b \gamma}_t$ in $\S$, so that
$\b \gamma_0=\epsilon$. The end point $\b \gamma_1$ lies in the fiber
$p^{-1}(e)$ which we identify with Z. We thus write $\b \gamma_1=k\in Z$.
Since the fibers are discrete, $k$ only depends on the homotopy class
$[\gamma_t]$ and we have $\partial_*([\gamma_t]):=k$. It is easy to see
that this map is in fact a homomorphisms from $Z$ to $Z$  which is hence
given by $\partial_*(n)=kn$. We do not yet know what the integer $k$ is.
Here we have:

\proclaim Lemma.
The homomorphism $\partial_*:$ is given by $n\mapsto mn$, where
$m$ is the monopole number.
\par

\noindent
{\bf Proof.} We must prove that the generator of
$Z\cong\pi_1\left(\G_{\infty}/\G_F\right)$ is mapped to $m$ times
the generator of $Z\cong
\G_F/\G_F^0\cong\pi_3(SO(3))$. To do this, we represent
the monopole configuration in the following form, into which each
configuration of monopole number $m\not =0$ may be smoothly deformed:
Take $m$ disjoint open 3-balls in ${\Re}^3$, $B_i$, $i=1,\dots ,m$,
and inside each of them a concentric and slightly smaller closed 3-ball
$B'_i$. The zeros of the Higgs field occur precisely at the $m$
centers of these balls. In the regions between the balls,
$S_i:=B_i-B'_i$, the Higgs field is radially pointing
with respect to the local centers. It points either in an outward
direction, in which case the monopole number is positive, or inward if
the monopole number is negative. We now take a real valued
$C^{\infty}$ function, $\rho$, that assumes the constant value one
outside all balls $B_i$ and zero inside the balls $B'_i$ and
only depends on the radial coordinate in each $S_i$ where it is
strictly monotonic. We then define a one parameter family of
maps $\b \gamma_t: {\Re}^3\rightarrow SO(3)$, $t\in[0,1]$, by
$$
{\b \gamma}_t(x):=
\exp\left\{2\pi t\rho(x){\phi^a(x)\over\Vert\phi(x)\Vert}T_a\right\}
\eqno{(10)}
$$
where $T_a$ denote the generators of $SO(3)$. This one parameter
family of maps just define the lifted curve $\b \gamma_t$
mentioned earlier. ${\b \gamma}_1$ maps the exterior
region ${\Re}^3-\cup_{i=1}^mB_i$ and the interior region
$\cup_{i=1}^m B'_i$ onto the identity. If we collapse each boundary
sphere $\partial B_i$ and $\partial B'$ of $S_i$ to a point, the resulting
3-sphere is wrapped twice onto $SO(3)$, since in $S_i$ opposite directions
with angles adding to $2\pi$ are mapped to the  same point in $SO(3)$.
The map ${\b \gamma}_1$ has thus degree $2m$ so that it represents the
element $m\in Z=\G_F/\G_F^0$, denoted by $[{\b \gamma}_1]$. On the
other hand, the maps ${\b \gamma}_t$ define a loop
$\gamma_t$ in $\G_{\infty}/\G_F$ whose homotopy class,
$[\gamma_t]$, generates $\pi_1(\G_{\infty}/\G_F)$.
{}From the discussion of the map $\partial_*$ we know that
$\partial_*[\gamma_t]=[{\b \gamma}_1]$. This proves the Lemma for
$m\not =0$. For $m=0$ we take only two concentric balls $B$ and $B'$
and the field configuration so that outside $B'$ the
Higgs field assumes a constant value in $S_H$. As before we take
a function $\rho$, build the 1-parameter family of maps (10)
and have $\partial_*[\gamma_t]=[{\b \gamma}_1]$. But now the map
${\b \gamma}_1$ has zero degree~$\bullet$

The exactness of (9) allows us to immediately infer from this Lemma
the triviality of $\pi_1(\S)$ and $\S/\S^0\cong Z_{|m|}$, for $m\not =0$.
For $m=0$ it implies $\pi_1(\S)\cong Z$ and $\S/\S^0\cong Z$. For
$m\not =0$ this means that the holonomy group of the bundle (8) is
the subgroup of integers in $\G_F/\G_F^0$ divisible by $m$. The
symmetry group is thus given by by the quotient
$$
\S\cong {Z\times \Re\over Z}
\eqno{(11)}
$$
where the group $Z$ in the denominator is generated by $(m,2\pi)$.
But this quotient is isomorphic to $Z_{|m|}\times \Re$, as the following
isomorphism explicitly shows (equivalence classes referring to the
$Z$-quotient are denoted by square brackets)
$$\eqalign{
& \theta:\, Z_{|m|}\times \Re  \rightarrow {Z\times \Re\over Z} \cr
& \theta(n,r) = [(n,{2\pi n\over m}+r)]                         \cr}
\eqno{(12)}
$$

\proclaim Theorem. The symmetry group $\S$ of a dyon configuration
with monopole number $m\not =0$ is isomorphic to $Z_{|m|}\times \Re$.
The interpretation of these factors  may be taken from (8) and (11).
\par

Due to the existence of discrete symmetries, the global $U(1)$
turns out to be neither identical to, nor a subgroup of the symmetry
group. Rather, the subgroup $mZ\subset Z$ (denoting the integers
divisible by $m$) extended $U(1)$ to become the universal covering
group $\Re$, which is non compact. This might be taken as topological
origin for the possibility of fractional charge in the quantum theory of
the model discussed here [2]. We emphasize how crucially this
depends on a careful separation of redundancy transformations (defined to be
those generated by the Gauss constraints) from among all allowed Gauge
transformations. For example, at the end of Ref. [2] it was remarked that --
in our notation -- generators of $\G_F/\G_F^0$ do not really
deserve to be called ``topologically non-trivial'' since they are still in
the identity component of $\G_{\infty}$. But in quantum
theory states are only required to be annihilated by the redundancy group,
so that the relevant topology is that of $\G_F$ and not $\G_{\infty}$.
If we treated the transformations in $\G_F/\G_F^0$ as redundancies,
we only would allow representations of $\S$ that restricted to the
trivial representation on $\G_F/\G_F^0$ and there would be no
fractional charges. If at all, restrictions on the representations
of the general symmetry group must be explained on physical
grounds. This could, for example, come about if one tries to
implement another group action on the state space as a symmetry.
It might then happen that the sectors for the irreducible
representations of $\G_F/G_F^0$ (labelled by $theta\in S^1$)
do not reduce the action of an additional symmetry group, except
for specific values of $\theta$. Well known is that $CP$ exchanges
the sectors $\theta$ and $-\theta$ so that an implementation
of $CP$ symmetry selects the $\theta$ values $0$ or $\pi$ [6].

Finally, we comment on our restriction of $\G_{\infty}$ to include only
asymptotically constant rotations. One could indeed envisage more
general choices, in which $\G_{\infty}/\G_F$ could in fact become an
infinite dimensional group. For example, one could require the maps
in $\G_{\infty}$ to extend to a 2-sphere compactification. In this case
${\Re}^3$ becomes the interior of a closed 3-ball whose 2-sphere boundary,
$S_{\infty}$, now corresponds to infinity. $\G_{\infty}/\G_F$ can then
be identified with a mapping space $M:S_{\infty}\rightarrow U(1)$ which forms
a group under pointwise multiplication. Our old choice would correspond to the
$U(1)$ subgroup of constant maps. However, given any fixed point
$x\in S_{\infty}$, we have a mapping $P:\,M\rightarrow U(1)$ defined by
evaluation at the point $x:\, \sigma\mapsto\sigma(x)$.
Now, this map in fact induces isomorphisms on the homotopy groups. To see
this, we also introduce $M_x=\{\sigma\in M\,/\,\sigma(x)=e\}$ where $e$
denotes the unit element in $U(1)$. For the space of base point preserving
maps $M_x$ it is indeed very easy to see that
$\pi_n(M_x)\cong\pi_{n+2}(U(1))\cong 1$. On the other hand, we have the
fibration
$$
\diag{M_x&\mapright{i}&M\cr
          &&\mapdown{P}\cr
          &&U(1)       \cr}
\eqno{(13)}
$$
whose associated exact sequence tells us that the maps
$P^{(n)}_*:\,\pi_n(M)\rightarrow\pi_n(U(1))$ are  isomorphisms for all $n\geq
0$.
This implies that the discussion following (8) indeed captures all the
nontrivial
topological implications presented by the discrete group group $\G_F/\G_F^0$.
It extends the $U(1)$ subgroup of asymptotically constant rotations about the
Higgs field in the way indicated by formula (11).

\beginsection{General Relativity}

In this last section we briefly indicate the applicability of the foregoing
to general relativity. Here we cannot attempt to account for all technicalities
so that some arguments are necessarily somewhat sketchy. Like gauge theories,
general relativity also deals with long ranged field configurations which are
usually taken to be asymptotically flat. The canonical variables are given by
a Riemannian metric, $g_{ik}$, and its conjugate momentum $\pi^{ik}$. Both
these tensor fields are defined on a 3-manifold $\Si$, the carrier space for
the Cauchy data. We assume $\Si$ to be without boundary and to possess only one
asymptotic region, that is, it contains a compact set outside which it is
homeomorphic to the complement of a closed ball in ${\Re}^3$.
After space-time is developed from the initial data, the momenta $\pi^{ik}$
can be expressed as a linear function of the extrinsic curvature
$K_{ik}$ of $\Si$ in space-time. The condition of asymptotic
flatness says that for large distances there exists a coordinate
chart in which the canonical variables have the following falloff
behavior:
$$\eqalign{
& g_{ik}(x)=\delta_{ik}+{a(n^i)\over r}+O(1/r) \cr
& \pi^{ik}(x)={b(n^i)\over r^2}+O(1/r^2)         \cr}
\eqno{(14)}
$$
where $n^i=x^i/r$. In addition, the functions $a$ and $b$ must be even
respectively odd under parity: $a(-n^i)=a(n^i)$ and $b(-n^i)=-b(n^i)$
(see [7] or [8])
Diffeomorphisms on $\Si$ must respect the asymptotic behavior (14).
The phase space function that generates infinitesimal such transformations
is given by ($\nabla$ denotes the Levi-Civita covariant derivative with
respect to the metric $g_{ik}$)
$$
I^{\xi}=2\int_{\Si}d^3x\,\nabla_i\xi_k\pi^{ik}
\eqno{(15)}
$$
In comparison, the diffeomorphism constraint reads
$$
D^k=-2\nabla_i\pi^{ik}=0
\eqno{(16)}
$$
so that
$$
\int_{\Si}d^3x\,\xi_kD^k=I^{\xi}-2\int_{S_{\infty}}n_i\pi^{ik}\xi_k
\eqno{(17)}
$$
Formulae (15)-(17) are just the analogs of (4)-(6) respectively.
The diffeomorphism constraints thus only generates the identity
component of the asymptotically trivial diffeomorphisms, for which the
surface term in (17) vanishes. In contrast, the identity component
of the diffeomorphisms compatible with the asymptotic conditions (14)
are generated by vector fields of the asymptotic form
$$
\xi_i=\varepsilon_{ijk}\rho^j x^k + \tau_i + O(1)
\eqno{(18)}
$$
where $\tau^i$ and $\rho^i$ are constant in the asymptotic
chart and represent rigid translations and rotations respectively
with respect to the asymptotically euclidean structure.
As before, the existence of discrete symmetries may cause the symmetry
group to be different from the euclidean group $E_3$. To see this,
we remark that it is sufficient to consider only the rotational
part and discard the translations within the allowed
diffeomorphisms since the mechanism described is entirely topological
in nature and thus insensitive to contractible parts of the group.
This is just as in the gauge theoretic case. In both cases it is
sufficient to retain only the maximal compact subgroup ($SO(3)$ here,
$U(1)$ there). This allows us to make use of the formal convenience
that all these diffeomorphisms extend to the 1-point compactification
$\Sib=\Si\cup\infty$. The allowed diffeomorphisms
then fix $\infty$ and reduce to $SO(3)$ rotations (with respect to
the preferred frame defined by the asymptotic chart) on the tangent
space at this point. We call this group, which is now our invariance
group, $\D_{\infty}$. The redundancy group, which is generated
by the diffeomorphism constraint, is given by the identity
component of those diffeomorphisms that not only fix $\infty$,
but also induce the identity map on the tangent space. We call
it $\D^0_F$, $F$ for frame-fixing and $0$ to denote the identity
component. It is clear that $\D_{\infty}/\D_F\cong SO(3)$ which gives
us the analog of (7):
$$
\diag{
\D_F&\mapright{i}&\D_{\infty}\cr
&&\mapdown{p}\cr
&&SO(3)\cr}
\eqno{(19)}
$$
Here the projection map $p$ is just the evaluation of the tangent
map at $\infty$. On the
other hand, the symmetry group $\S$ is defined by
$\S=\D_{\infty}/\D^0_F$. Again we wish to know how the discrete
normal subgroup $\D_F/\D^0_F$ combines topologically with $SO(3)$
to form $\S$. In full analogy to (8) we have
$$
\diag{
\D_F/\D^0_F&\mapright{}&\D_{\infty}/\D^0_F\cong \S\cr
&&\mapdown{p}\qquad \cr
&&SO(3)\qquad\cr}
\eqno{(20)}
$$
In distinction to (8), we did not indicate what the group
$\D_F/\D^0_F$ is, since this depends on the topology of the underlying
3-manifold $\Sib$. We can nevertheless continue the analogy, and
write down the final piece of the exact sequence associated with
(20)
$$
\diag{
1&\maprights{}&\pi_1(\S)&\maprights{p_*}&Z_2&\maprights{\partial_*}&
\D_F/\D^0_F&\maprights{i_*}&\S/\S^0&\maprights{}&1\cr
&&&&\mapdowns{iso}&&&&&&\cr
&&&&\pi_1(SO(3))&&&&&&\cr}
\eqno{(21)}
$$
With the obvious adaptations we can almost literally transfer
the discussion following Eqn. (9) for the map $\partial_*$.
A more geometric description that is analogous to the
the one surrounding formula (10) goes as follows:
Pick a loop $\gamma_t$, $t\in[0,1]$, in $SO(3)$ whose homotopy
class $[\gamma_t]$ generates $\pi_1(SO(3))\cong Z_2$. Take a closed
embedded ball $B\subset\Sib$ centered at $\infty$ with standard
spherical polar coordinates $(r,\theta ,\varphi)$ with
$0\leq r\leq 2$, i.e., $r=2$ corresponds to the boundary $\partial B$.
Let the ball $r\leq 1$ be called $B'$. Let
further $\rho$ be a smooth monotonic function
$\Re\rightarrow \Re$ so that  $\rho(x)=0$ for $x\leq 0$ and
$\rho(x)=1$ for $x\geq 1$. Define a curve ${\b \gamma}_t$ in
$\D_{\infty}$ in the following way: outside $B$ $\b \gamma_t$
is the identity and inside $B$ it is defined by
${\b \gamma}_t(r,\theta, \varphi):=(r,\theta, \varphi +
2\pi t\rho(2-r))$. This defines a lift of $\gamma_t$ into
$\D_{\infty}$. The end point, ${\b \gamma}_1$, corresponds to
the identity map inside $B'$ and thus defines an element in
$\D_F$. It is called a rotation parallel to the spheres
$\partial B$ and $\partial B'$, or, since $\infty\in B'$, simply
a `rotation at infinity'. We denote by $[\b \gamma_1]\in \D_F/\D_F^0$
its mapping class within in $\D_F$. The map $\partial_*$ is now
defined by $\partial_*([\gamma_t])=[{\b \gamma}_1]$. This map is
well defined and in fact a homomorphism.

Comparing (9) and (21) we see that now the third entry in the
sequence is $Z_2$ rather than $Z$ which leaves us with only two
possibilities:

\noindent
Case I. $\pi_1(\S)\cong Z_2$ and $\hbox{Image}(\partial_*)=\{1\}$
(the trivial group). In this case the bundle (20) is trivial.
Geometrically this means that the rotation at infinity is an
element of $D_F^0$, i.e., in the identity component. The
symmetry group is then just given by
$$
\S\cong SO(3)\times \left\{\D_F/\D^0_F\right\}
\eqno{(22)}
$$

\noindent
Case II. $\pi_1(\S)\cong \{1\}$ and $\hbox{Image}(\partial_*)
\cong Z_2$. In this case the bundle is non-trivial with $Z_2$
holonomy. Geometrically this means that the rotation at infinity
is not in the identity component $\D_F^0$.  We have
$$
\S\cong {SU(2)\times\left\{\D_F/\D^0_F\right\}\over Z_2}
\eqno{(23)}
$$
where the $Z_2$ in the denominator is `diagonal', that is,
it is generated by $(-1,-1)$ where the left $-1$ generates
the center in $SU(2)$ and the right $-1$ some (in fact
central) $Z_2$ subgroup of $\D_F/\D^0_F$ which is generated
by a rotation at infinity.

Now, whether case I or II is realized depends purely on the
topology of the manifold $\Sib$. Both possibilities occur,
and it is established for all known 3-manifolds under which case
they fall. The working result is in fact easy to communicate:
Let $\Sib$ be any of the presently known compact orientable
3-manifolds. It falls under case I, if and
only if it is the connected sum of handles ($S^1\times S^2$) and
lens spaces ($L(p,q)$) (the 3-sphere is included here as the
space $L(1,1))$. For more details we refer to [9] and references
therein. The interesting thing about
case II is that the symmetry group simply does not have an $SO(3)$
rotational subgroup. Manifolds in this class may therefore be
termed {\it spinorial}.
It has been argued that these manifolds could give rise to
odd half-integer angular momentum states in quantum gravity
which would thus be stabilized against decay in pure gravity
[10] (see also [11] for a more recent survey).
A general investigation on the structure of the group
$\D_F/\D_F^0$ for arbitrary 3-manifolds $\Sib$ will appear
elsewhere [12].

\beginsection{Summary}

We started with the observation that in gauge or diffeomorphism
invariant theories, not all gauge transformations correspond to
redundancies in the presence of long ranged configurations.
Rather, it is the Gauss constraint that declares some of the
formally present degrees of freedom to be physically non existent.
But it only generates the identity component of asymptotically
trivial transformations, leaving out the long ranging ones with
preserve the asymptotic structure imposed by boundary conditions as
well as those not in the identity component of the asymptotically
trivial ones. These should be considered as proper physical
symmetries which act on physically existing
degrees of freedom. For example, asymptotic $U(1)$ gauge rotations
of a dyon with non vanishing electric charge, or asymptotic spatial
rotations of a black hole with non vanishing angular momentum cost
physical action. In order to establish the structure of these symmetry
groups one needs to take
into account the asymptotically trivial transformations not connected
to the identity.  These generally do not form a
factor of the full symmetry group, but are rather positioned in
a topologically non trivial way so as to reduce the number of
connected components. In our examples this was described by
formulae (11) and (23). As a result, the group $U(1)$ was turned
into its non compact universal cover $\Re$. In gravity, the change
from the $SO(3)$ of spatial rotations to its universal
cover $SU(2)$ was only a possibility (although quite generic),
depending on the topology of the underlying 3-manifold $\Sib$.

\bigskip

\noindent{\bf Acknowledgement}
I like to thank the Center of Gravitational Physics and Geometry
at the Pennsylvania State University for its hospitality. This
work was supported by NSF grant PHY-9396246 A001.
\beginsection{References}
\parindent=0pt

\item{[1]}  J. Marsden, A. Weinstein. {\it Rep. Math. Phys.} {\bf 5},
            121-130 (1974). Also, R. Abraham, J. Marsden, Foundations
            of Mechanics, 2nd edition, chapter 4. The Benjamin
            Cummings Publishing Company (1982).
\item{[2]}  E. Witten. {\it Phys. Lett.} {\bf 86B}, 283-287 (1979).
\item{[3]}  C. Aneziris, A.P. Balachandran, M. Bourdeau, S. Jo,
            R.D. Sorkin, T.R. Ramadas. {\it Int. Jour. Mod. Phys. A}
            {\bf 20}, 5459-5510 (1989)
\item{[4]}  B. Julia, A. Zee. {\it Phys. Rev. D} {\bf 11}, 2227-2232
            (1975)
\item{[5]}  P. Goddard, D. Olive. {\it Rep. Prog. Phys.} {\bf 41},
            1357-1437 (1978)
\item{[6]}  A. Ashtekar, A.P. Balachandran, S. Jo. {\it Int. Jour.
            Mod. Phys. A} {\bf 6}, 1493-1514 (1989)
\item{[7]}  T. Regge, C. Teitelboim. {\it Ann. Phys. (NY)} {\bf 88},
            286-318 (1974)
\item{[8]}  R. Beig, N. \'O Murchadha. {\it Ann. Phys. (NY)}
            {\bf 174}, 463-498 (1986)
\item{[9]}  D. Giulini. {\it Int. Jour. Theo. Phys.} {\bf 33},
            913-930 (1994)
\item{[10]}  J. Friedman, R.D. Sorkin. {\it Phys. Rev. Lett} {\bf 44},
            1100-1103 (1980)
\item{[11]} J. Friedman. Space Time Topology and Quantum Gravity.
            In: Conceptual Problems in Quantum Gravity,
            Proceedings of the 1988 Osgood Hill Conference, Ed.
            A. Ashtekar and J. Stachel. Einstein Studies,
            Vol. 2, Birkh\"auser, Boston, Basel, Berlin (1991).
\item{[12]} D. Giulini. Discrete Symmetries in General Relativity.
            In preparation.

\end